\documentclass[11pt,twoside]{article}

\usepackage{asp2006}
\usepackage{epsf}
\usepackage{psfig}
\usepackage{lscape}

\begin{document}

\title{Jet Interactions with the Hot Halos of Clusters and Galaxies}   

\author{B.R. McNamara$^{1,3}$, L. B\^{\i}rzan$^{2}$, D.A. Rafferty$^2$, P.E.J. Nulsen$^3$, C. Carilli$^{4}$, M.W. Wise$^{5}$}   
\affil{1) Dept. of Physics \& Astronomy, University of Waterloo, 200 University Avenue West, Waterloo, ONT, Canada N2L 3G1 \\
2) Dept. of Physics \& Astronomy, Ohio University, Clippinger Labs, 
Athens, OH, 45701 U.S.A.\\ 
3) Harvard-Smithsonian Center for Astrophysics, 60 Garden St., Cambridge, MA,
02138 U.S.A.\\ 
4) NRAO, P.O. Box 0, 1003 Lopezville Rd., Socorro, NM, 87801-0387 U.S.A.\\ 
5) Astronomical Institute ``Anton Pannekoek'', Kruislaan 403, 1098 SJ Amsterdam, the Netherlands}    

\begin{abstract} 
X-ray observations of cavities and shock 
fronts produced by jets streaming through 
hot halos have significantly advanced
our understanding of the energetics and
dynamics of extragalactic radio sources.
Radio sources at the centers of clusters 
have dynamical ages between ten and several
hundred million years.  They liberate between $10^{58-62}$ erg 
per outburst, which is enough energy to regulate cooling of hot halos
from galaxies to the richest clusters.
Jet power scales approximately with the radio synchrotron luminosity
to the one half power. However, the synchrotron efficiency 
varies widely from nearly unity to one 
part in 10,000, such that relatively feeble radio source can have
quasar-like mechanical power.  
The synchrotron ages of cluster radio sources are decoupled from
their dynamical ages, which tend to be factors of several to orders
of magnitude older.  Magnetic fields and particles in the lobes tend 
to be out of equipartition. The lobes may be maintained 
by heavy particles (e.g., protons), low energy electrons,
a hot, diffuse thermal gas, or possibly magnetic (Poynting)
stresses.  Sensitive X-ray images of 
shock fronts and cavities can be used to study the dynamics
of extragalactic radio sources. 

\end{abstract}

\section{X-ray Cavities in Hot Halos}

Cavities embedded in hot X-ray halos gauge the energy
and power output of recent AGN activity in central galaxies.
\footnote{See \citet{pf06} and \citet{mn07}  for
recent reviews of this topic, including  discussions of
theory and simulation which has been omitted from this paper.}
Cavity systems are found in nearly three dozen 
clusters \citep[]{brm04,df04,r07}
and a similar number in giant elliptical galaxies \citep{nuls06}.
Like the extragalactic jets that created them, cavities
are usually found in pairs.  Their surface brightness decrements
of several tens of percent are consistent with being nearly devoid
of thermal gas at ambient conditions.  Most are filled 
with radio emission, and some
are connected to the nucleus through synchrotron jets and tunnels
in the hot gas \citep[Fig. 3]{csb05, wise07}.  
Many have bright rims composed of cool, dense, gas,
ruling-out their association with shock fronts. Most cavities are
apparently in pressure balance or are expanding or contracting
slowly with respect to their surroundings.  In many instances, buoyancy
forces are driving them outward.  Then their locations with respect
to their origin, generally taken to be the nucleus, provide a
measure of their dynamical ages, which are generally $\sim 10^7-10^8$ yr.
Cavity diameters in clusters are typically 20 kpc but can be as large
as 200 kpc. In galaxies they are typically a few kpc in diameter.  
Assuming they form near the nucleus, cavities live long enough to
rise beyond their own diameters before breaking apart or fading
into the background. Thus cavities must be 
pressurized and their rims stabilized by motion through
the atmosphere \citep{ps06}, 
plasma viscosity \citep{rey05}, or magnetic fields draping their surfaces \citep{deyoung03, jd05, lyut06}.  In some cases, the
cavities appear to be breaking apart where radio plasma is
leaking into the ICM \citep{bsm01} or into outer ``ghost'' 
cavities \citep{fab00}.  

\section{Cavity Energetics}

X-ray observations provide accurate measurements of the run of
gas temperature and
density in the halo, and thus its pressure. The
projected sizes of cavities and hence their volumes then provide
the work, $pV$, required to inflate the cavity against the surrounding
gas pressure \citep{mcn00}.  Including the energy of the plasma filling them,
their total enthalpy can be expressed as
$H=\gamma pV/(\gamma -1)$.  The (unknown) ratio of specific heats
of the filling plasma, $\gamma$, 
implies a total enthalpy between $2.5pV-4pV$ per cavity \citep{churaz02}. 
Using the rise
time, $t_{\rm rise}$, the average jet power 
$P_{\rm jet}=H/t_{\rm rise}$ can be found.    

Jet power estimated in this manner spans eight orders of magnitude
from $10^{38}~{\rm erg~s^{-1}}$ in gE galaxies and 
up to $10^{46}~{\rm erg~s^{-1}}$
in rich clusters.  Typical values in rich clusters are 
$10^{42}~{\rm erg~s^{-1}}$ to $10^{44}~{\rm erg~s^{-1}}$, rivaling
and sometimes exceeding
the total X-ray luminosity of cluster cores.  Several clusters
and galaxies have weak shock fronts surrounding their radio
sources and cavities \citep{fab03, mcn05, nuls05, 
forman05}, two of which are shown in Fig. 3.  
Mildly transonic Mach numbers $\sim 1.2-1.6$
are found in clusters, while stronger shocks are found in lower
mass giant elliptical systems \citep{kraft03}.
The outburst energy estimated independently from
shocks lies reassuringly within small multiples of the 
cavity power. Thus, jet power estimated by cavity dynamics
provides a reliable lower limit to the total mechanical energy dissipated
by radio jets.

\section{Feedback in Galaxies \& Clusters}

Cavity systems are found preferentially in galaxies and clusters
whose hot halos have cooling times $\le 5\times 10^8~{\rm yr}$
\citep{df06}.
In such systems gas is expected to cool at rates of 
$\le 1~ {\rm M_\odot ~yr^{-1}}$
in gEs and $100-1000~ {\rm M_\odot ~yr^{-1}}$ in clusters.  However, high-
and moderate-resolution X-ray spectroscopy made with the Chandra
and XMM-Newton X-ray observatories failed to detect the characteristic
emission below 1 keV from gas cooling at these rates \citep{pf06}.  This
implies that gas is either cooling non-radiatively into
an unseen form, or the gas is maintained at high temperatures
by a powerful and persistent energy source.  The former idea has
been essentially ruled out by sensitive searches for such a repository
carried out during the past three decades. However, several viable
energy sources have been explored, including 
thermal conduction from the hot outer
halos, AGN, mergers, cosmic rays, and other mechanisms.  All of these
mechanisms may be operating in one system or another to a
greater or lesser degree. But AGN feedback appears to be
operational and energetically important in most halos where
heating is required to offset cooling.  

There are several lines of supporting evidence.
First, the apparent ubiquity of supermassive black holes (SMBH) in 
the nuclei of bulges \citep{magor98}
and the high efficiency with which SMBHs
convert gravitational binding energy into mechanical energy and 
heat make them an ideal and nearly universal source of energy in
hot halos.  

Second, nearly 70\% of cooling flow
clusters with high cooling rates have active cavity systems and or radio sources
indicating they are ``on'' much of the time \citep{df06}. 
Roughly 20\%
of gEs harbor detectable cavity systems \citep{nuls06}.
AGN liberate on average the 
right amount of energy to balance radiative
losses over many orders of magnitude of power in isolated
gE galaxies to the richest clusters.  Roughly half of
cluster cavity systems are underpowered and those preferentially
experience star formation at rates of several to several
tens of solar masses per year \citep{r06,r07}.  In contrast,
all of the gE cavity systems are overpowered relative to their
cooling luminosities and should be able to quench star formation
at late time, as is observed.  A picture of this feedback sequence
adapted from the heating and cooling diagrams of 
\citet{brm04}, \citet{df06}, and \citet{nuls06} but painted with a broad brush
is given in Fig. 1. There are individual exceptions to the
partitioning in this diagram, but it provides a reasonably good
characterization.  

Third, feedback
from a central source appears necessary to maintain the
high gas densities and short central cooling times \citep{vf04}
without allowing catastrophic cooling.  Finally, cluster halos show
entropy cores indicating a history of central energy injection
that is consistent with the current rate of cavity 
injection \citep{vd05}.  We still do not understand
how AGN heat the gas, although there are many ideas 
\citep[see][for further discussion]{mn07}.  This
problem will hopefully sort itself out over the next several years. 

\begin{figure}[ht]
\begin{center}
\hbox{
\hspace{1.0cm}
\psfig{figure=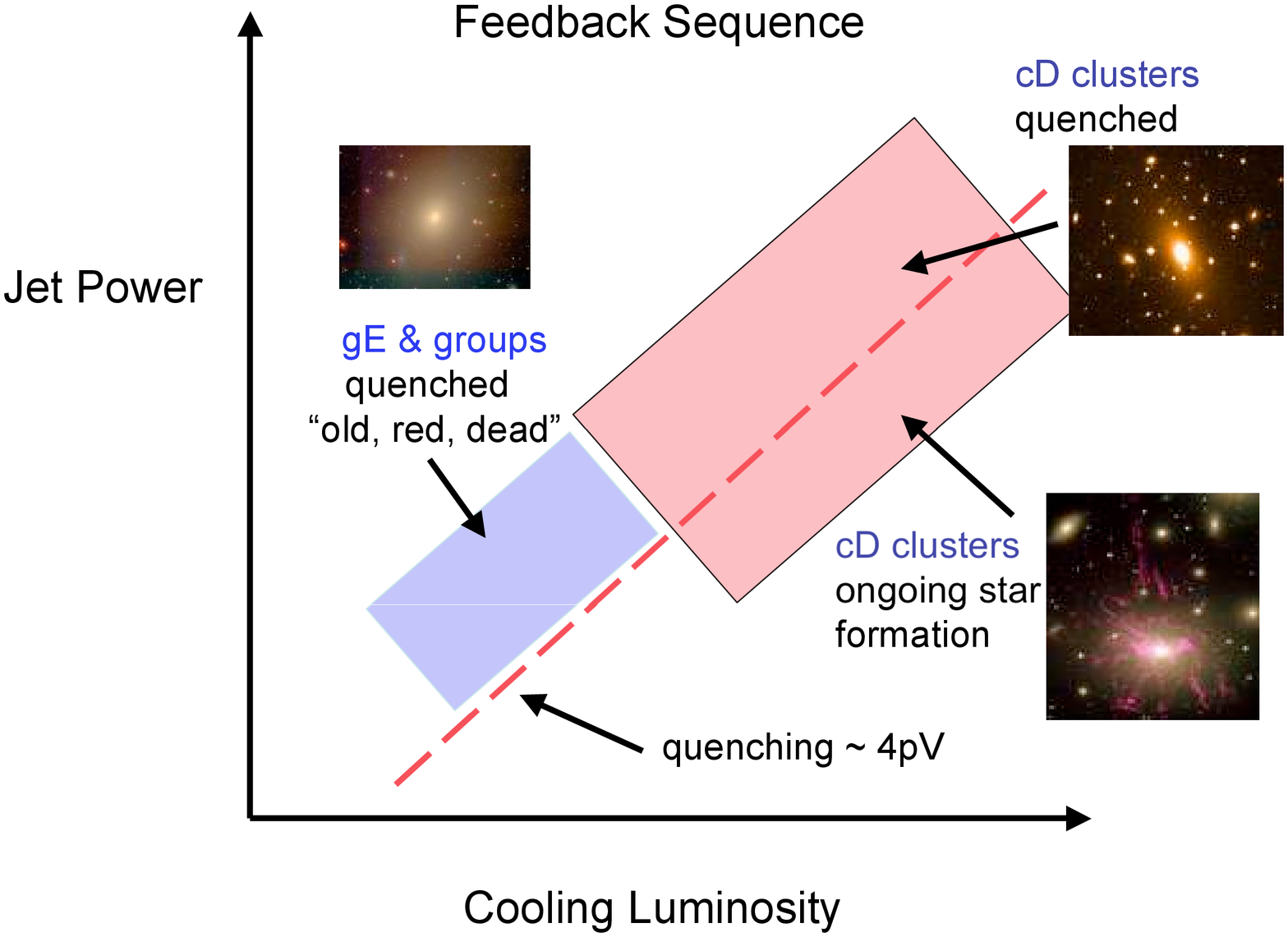,height=8.0cm,width=10.0cm,angle=-90}
}
\vspace{-1.0cm}
\end{center}
\caption{\footnotesize
Depiction of conjectured AGN feedback sequence that runs from
giant elliptical halos (lower left) to rich clusters
(upper right). Objects lying above the ``quenching'' line
are energetically able to maintain their halos at X-ray
temperatures at the present time. }
\label{fig1}
\end{figure}

\section{General Consequences of Feedback in Clusters}

AGN feedback lies at the heart of several problems in 
galaxy formation, including the exponential decline rather
than the expected power law behavior of the bright
end of the galaxy luminosity function \citep{benson03}, 
the existence of
the Magorrian (1998) relation (the correlation between SMBH mass
and bulge mass), and ``cosmic downsizing,'' the anti-hierarchical tendency
for the most massivive galaxies to lie dormant while low
mass systems burgeon with star formation at late time \citep{jun05}.   
Each may be a consequence of feedback of one form or another, although
AGN surely dominate in massive systems. 
Cooling flows are
interesting and particulary important in this respect.
They are among the few if only systems where 
cooling, star formation, and feedback in large
halos are clearly manifest such that the theory of feedback
and jet interaction can be tested in detail.
The spate of discoveries of cavities and shock fronts in hot halos
over the past eight years has provided much of the impetus for 
new, feedback-based models of galaxy formation and evolution.  
Finally, cluster-scale outbursts like those in Fig. 3 may contribute 
significantly to the excess entropy (``preheating'') in groups
and clusters \citep{vd05}.

\begin{figure}[ht]
\begin{center}
\hbox{
\hspace{1.0cm}
\psfig{figure=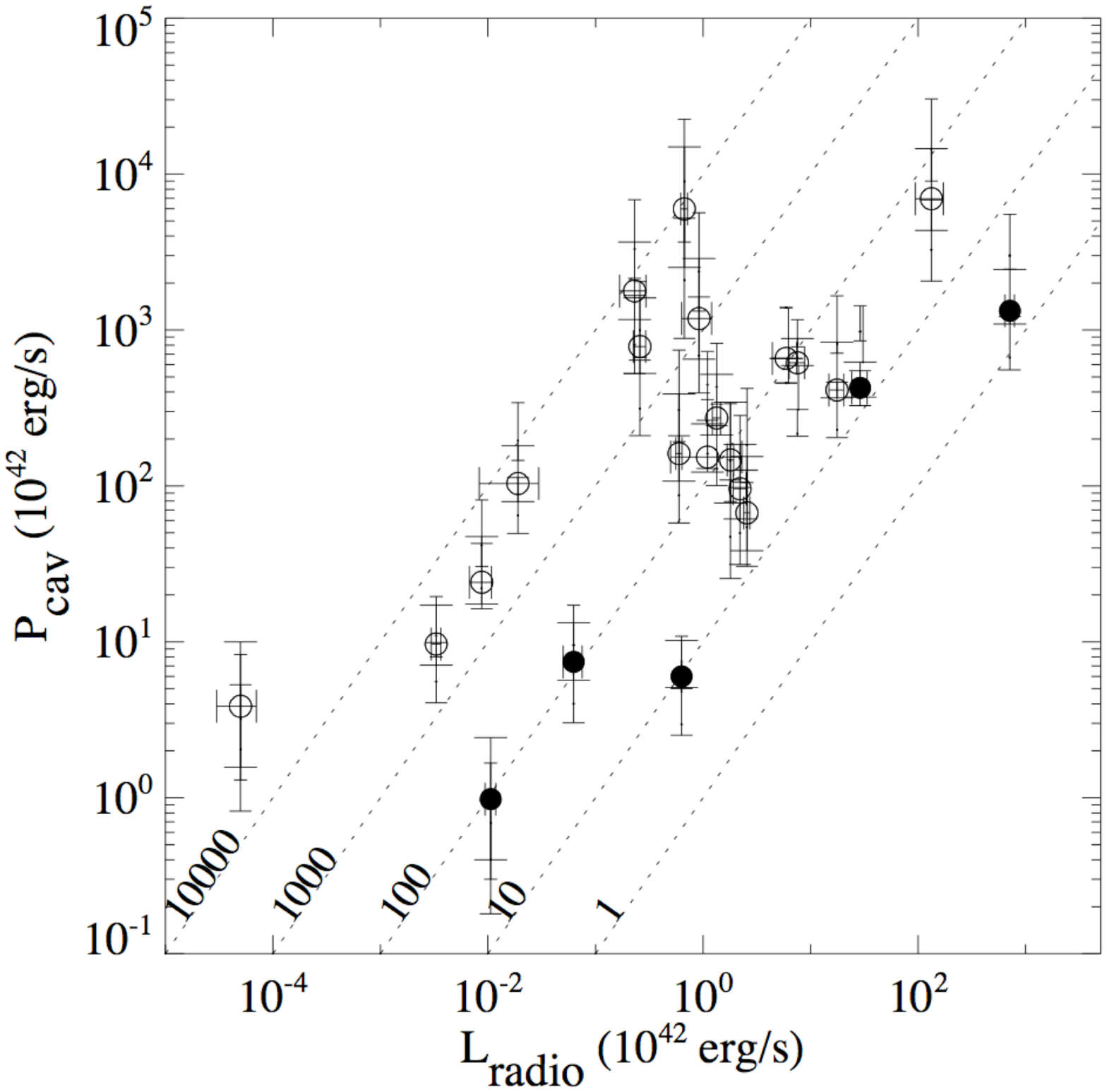,height=10.0cm,width=12.0cm,angle=90}
}
\vspace{-1.0cm}
\end{center}
\caption{\footnotesize
Cavity power versus bolometric synnchrotron power in the cavities
alone from
B\^{\i}rzan et al. (2007). Diagonal lines represent constant
values of $\epsilon$. Solid points are cavity systems that have
significant 8 GHz emission. }
\label{fig2}
\end{figure}

\section{Using Cavities to Gauge AGN Mechanical Power }

\subsection{Synchrotron Radiation Efficiencies of Radio Sources}

Cavity systems and their associated shock fronts provide a reliable
probe of the energetics and dynamics of radio outbursts.
It is commonly held that the energy in a radio source is
partitioned between particles and magnetic fields, 
$E_{\rm tot}=E_{\rm B}+E_{\rm p}$.
Radio observations of jets and lobes measure their synchrotron radiation,
revealing the existence of relativistic electrons
and magnetic fields, but not their total power. 
It has long been thought that the total power is dominated
by the mechanical work driving their expansion \citep{scheuer74}.
However, prior to Chandra, the total energy 
had to be inferred indirectly (e.g., \citet{will99}.  
A measure of the ratio of total jet power to
synchrotron power is given by the 
radiative inefficiency ratio $\epsilon=P_{\rm jet}/L_{\rm syn}$. 
The magnitude of this ratio is often assumed to lie between a few
and 100, with $\epsilon = 10$ taken as a reasonable estimate for
many sources.  

An early systematic cavity
study by \citet{brm04} based on radio data taken from
the literature found that jet (cavity) power scales 
with radio synchroton luminosity to the one half power, but with a very large
scatter in $\epsilon$.  B\^{\i}rzan and
colleagues have recently updated this work with new VLA observations
of 23 cavity systems at three or more frequencies, shown in Fig. 2.
Her work has confirmed the reality of the trend and
the large variations in $\epsilon$.  The median 
$\epsilon_{\rm med}\simeq 120$ for the lobes alone, but the average 
$\epsilon_{\rm av}\simeq 4700$ is dramatically larger.  The range
includes values of approximately unity for the youngest sources 
and more than 10,000 in the
case of the cavity system in the HCG 62 group.  Note that these
figures are underestimates as they do not include the energy
released by shocks, which can be factors of several larger.
$\epsilon$ is almost surely not constant in time, and is expected
to increase due to adiabatic expansion and synchrotron aging.
But these factors may not account for such
large variations, and there may be substantial intrinsic variations
in $\epsilon$.

This result is consequential in several respects.  It shows 
quantitatively that FR I-type AGN are on average 
strongly dominated by mechanical energy.
In some cases, feeble FR-I sources with dynamical 
ages of a few tens of Myr are as mechanically powerful as luminous quasars.  
Thus, feedback from modest radio sources can have a dramatic effect
on their environment by driving outflows, stifling star formation,
and quenching cooling flows.  The  unexpectedly large
variation in $\epsilon$ is one of the principal reasons 
that AGN have only recently become widely recognized
as a viable quenching mechanism for cooling flows.  Finally, it
is unfortunate news for galaxy formation modelers and 
observers seeking a simple 
prescription for feedback based on radio surveys of galaxies.  There
is no simple conversion between jet power and synchrotron
luminosity. The situation is more complex than we might have hoped.

\subsection{Jet \& Lobe Contents}

The dependencies in the energy equation can be expressed as:
$$ E_{\rm tot} = E_{\rm B}+E_{\rm p}\propto {B^2\over 8\pi}\Phi V + A(1+k)L_{\rm syn}B^{-3/2},$$
where $B$ is the magnetic field strength in the lobes, $V$ is the lobe
volume, $L_{\rm syn}$ is the bolometric synchrotron luminosity,
$\Phi$ is the field filling factor, and $k$ is the ratio of energy
in particles to that in electrons.  Additional terms may be
required to account for additional sources of pressure.  A 
hot, diffuse thermal plasma is an obvious example.

X-ray observations provide several constraints. $E_{\rm tot}\sim pV$.
$\Phi$ is inferred to be close to unity or cavities, which are
difficult to detect under ideal circumstances, would be invisible 
were they not nearly swept of ambient gas.  Cavity volumes are estimated
from their projected sizes.
Only $L_{\rm syn}$ is measured with radio observations at several frequencies 
using instruments such as the Very Large Array.  
This leaves $k$ and $B$ as free parameters that can be inferred 
statistically from large samples of cavities under the (reasonable) assumption
of pressure balance with their surroundings.

Analyses along these lines were
presented by \citet{df04} and \citet{dft05}.
Using radio data taken from the literature, they found a large 
variation in $k$ such that $1 \le k\le 1000$. In most systems
$k> 1$ indicating that the energy is dominated by heavy particles.
\citet{deyoung03}
reached a similar conclusion by treating observed jets 
as pipe-like conduits through which 
all of the energy in cavities must pass in a timescale comparable
to the synchrotron and dynamical ages.  He found the required energy 
desities within the jets must be so much larger than the surrounding
gas pressures that jets would quickly
decollimate unless the jet fluid
is composed of heavy particles (eg., protons) with an anisotropic
pressure field.  Otherwise, the jets must be actively collimated
by magnetic fields, for example, which has its own complications, or
perhaps they are dominated by electromagnetic flux 
\citep[e.g.,][this conference]{blan76, nld07}.
   
Using the new VLA sample, \citet{birz07}
found a similarly large variation in $k$, but in general
$k>>1$ and $B\le 50 \mu$G inside the lobes.  $B$ and $k$ vary widely
from system to system.  How much of 
this is intrinsic
or is a consquence of degeneracy between $B$ and $k$ and other assumptions
is less clear.  There are indications that
$k$ may be somewhat time dependent due to synchrotron aging and
adiabatic expansion.  But the findings of large $k$ appear to
be robust.  It may be difficult to get around the conclusion that
that FR I radio sources in clusters are heavy.  It is less clear
that they are born that way or become heavy by
entraining ambient material as they advance.  

Other possibilities include a dominant population of 
mildy relativistic, low energy electrons \citep{will99}.
The large $k$s may be a consequence of a missing third term accounting for
additional pressure support by a hot thermal plasma.
Limits on such a plasma indicate gas temperatures $T \ge 15-20$ 
keV \citep[see][for references and a discussion]{mn07}. 
Finally, it was pointed out in this conference by Hui Li and his
collaborators that their models of  current-carying Poynting 
jets moving in pressure balance with the surrounding gas are able
to reproduce the distribution of cavity sizes with nuclear distance found
by \citet{r06} and \citet{wise07}.
 
\begin{figure}[ht]
\begin{center}
\hbox{
\hspace{-1.0cm}
\psfig{figure=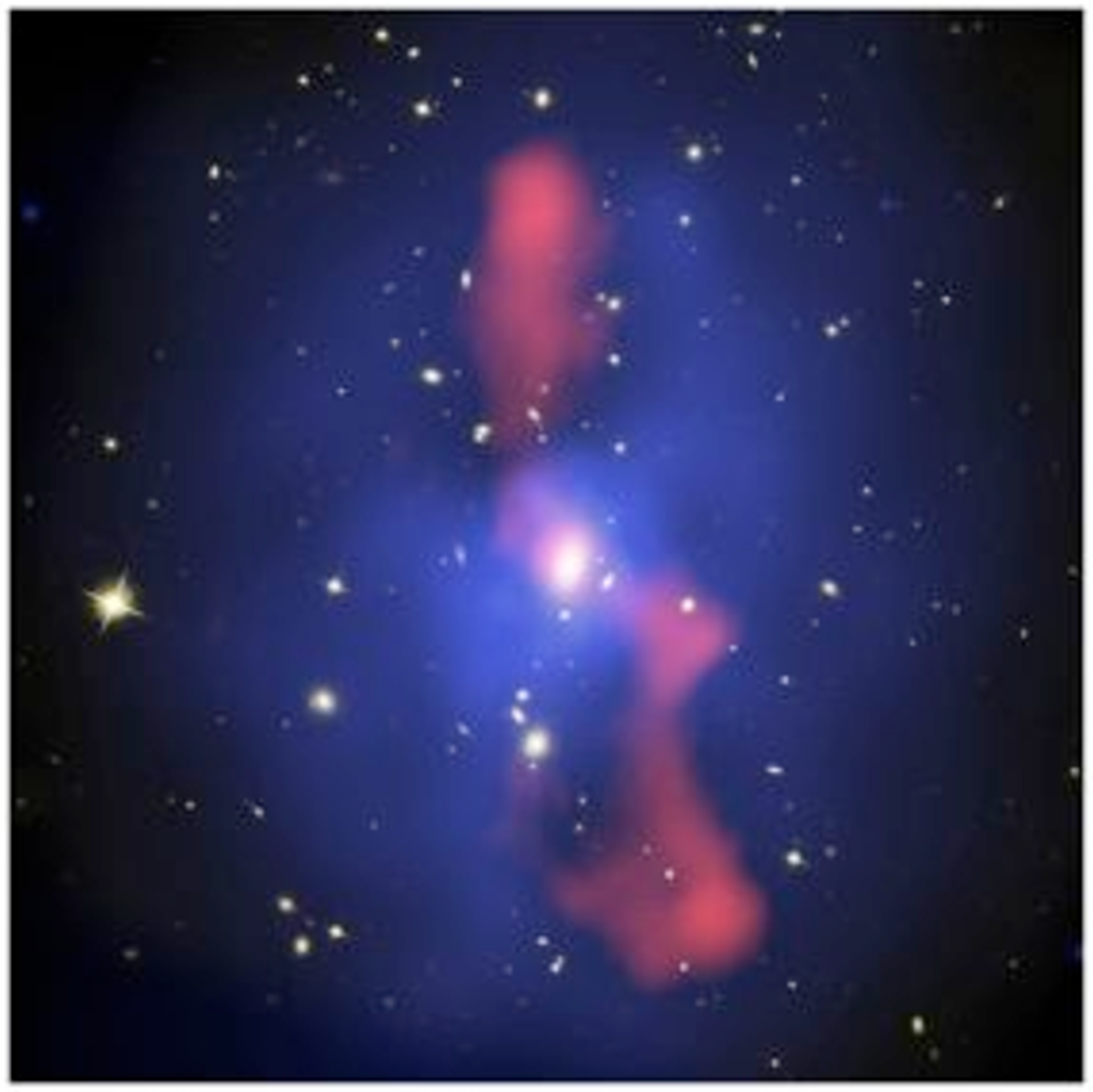,height=10.0cm,width=8.0cm,angle=0}
\psfig{figure=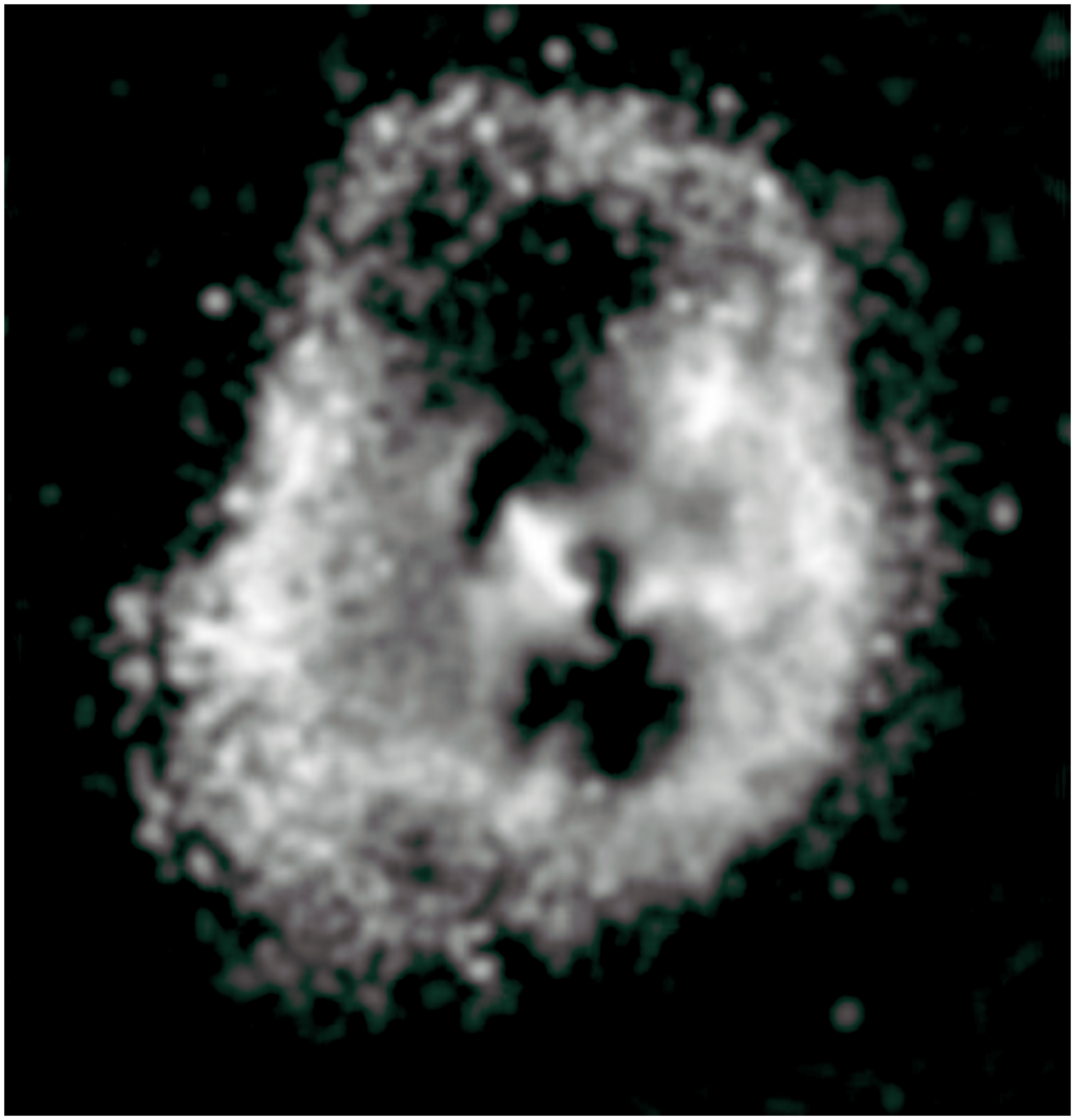,height=10.0cm,width=8.0cm,angle=0}
}
\vspace{-1.0cm}
\end{center}
\caption{\footnotesize
Two cluster-scale AGN outbursts with powers exceeding
$10^{61}$ erg. MS0735.6+7421 is shown at left \citep{mcn05}
and Hydra A \citep{wise07} is  shown at right.  In MS0735,
X-rays are in grey, 327 MHz emission \citep{birz07}
is in white as are the galaxies.  
The twin cavities are each roughly 200 kpc (one arcminute)
across. This figure is available in color at the public websites
of the Chandra X-ray and Hubble Space Telescope Observatories. 
The central $\simeq 12$ arcmmin (750 kpc) of Hydra A (right) in X-rays
is filled with
a complex system of cavities. The central cavities of \citet{mcn00}
each roughly 20 kpc in diameter are
dwarfed by the outer cavities.  The largest cavity to the north
\citep[cavity E of][]{wise07} lies 226 kpc from the nucleus
and has a diameter of roughly 200 kpc. The entire cavity system
is filled with 327 MHz radio emission \citep{lane04}.
The cavity systems in both clusters are enveloped by weak shock fronts. }
\label{fig3}
\end{figure}

\subsection{ Equipartition }

Adopting the well justified assumption that cavities
are in pressure balance with their surroundings, one can evaluate
whether the pressure support can be supplied by a plasma
in equipartition between electrons and magnetic field.  
Results for individual objects, such as Abell 2052 \citep{bsm01}, 
have shown that the internal lobe pressures must 
exceed the equipartition values in order to maintain pressure 
balance with the ambient gas.
Dunn's (2005) analysis 
and B\^{\i}rzan's new analysis of 19 systems (for $k < 100$) 
both find this to be true in general.  Equipartition pressures lie a factor 
of 10 or more below the values required to maintain pressure 
balance, apart from a few systems, such as Cygnus A and
the inner cavities of Hydra A.

\section{Jet and Lobe Dynamics}

B\^{\i}rzan's radio analysis permits a comparison of
the cavity buoyancy (dynamical) ages to their synchrotron
ages, providing an independent check on both quantities.
The ages agree in several instances.  However, by and large
the cavity ages exceed the synchrotron ages by factors of
several to several tens.  These differences probably reflect
to some degree weaknesses in the foundations
of both age estimates.  A strong constraint
can be placed on the relative ages by the presence
or absence of strong shocks in the hot halos.  In order to bring the
dynamical ages into agreement with the synchrotron ages
outward velocities that are factors of several
to several tens larger than their buoyancy speeds are implied.
These speeds are highly
supersonic.  Such fast moving cavities would create strong shocks with
Mach numbers much higher than those observed.  Overall,
the data suggest that the radiation ages and the dynamical
ages of these systems are decoupled.
 
Another check on the buoyancy ages can be made by comparing
them to the entirely independent ages of the shock fronts that
sometimes accompany them.  This check has been performed
in MS0735 and Hydra A, and in both cases the ages agree
to within a factor of two.  Hydra A is interesting in that
the buoyancy age of the large outer cavity to the north
shown in Fig. 3 \citep[cavity E of][]{wise07} is 220 Myr 
compared to the shock age of 140 Myr.  Although they
are in formal agreement, taken at face value
these ages are inconsistent with each other if 
the cavities are currently driving the shock.  The implications
are that either the cavities are still being driven by the
jets and thus the buoyancy age overestimates the true age, or
the cavities form far from the nucleus.
It is possible, though less likely, that the shock formed later 
and has just overrun the outer cavities.

The ratio of energy in the cavities to that in the shock fronts can in
principle indicate whether the AGN is ramping up or down.
Both in Hydra A and MS0735 this ratio is only modestly less
than unity, indicating that the jets are still powering
the cavities.  The situation
in Hydra A is complex (see Fig. 3), and thus a clear picture has
yet to emerge.  The point here is that deep Chandra observations
combined with multiple frequency radio observations
of cavity and shock systems have the potential
to provide a detailed
view of the dynamics of extragalactic radio sources.



\acknowledgements 
BRM thanks the LOC and SOC for organizing a splendid meeting and
for the invitation to speak.  This research was funded in part 
by NASA Long Term Space Astrophysics Grant NAG4-11025 and by
a grant from the Natural Sciences and Engineering Research 
Council of Canada.


\end{document}